# Friction tunable electrostatic clutch with low driving voltage for kinesthetic haptic feedback

Jongseok Nam[1], Jihyeong Ma[1], Nak Hyeong Lee[1], and Ki-Uk Kyung[1]

[1] *Department of Mechanical Engineering, Korea Advanced Institute of Science and Technology,*

*Daejeon, Republic of Korea*

*(Email:kyungku@kaist.ac.kr)*

**Abstract** --- As interest in Virtual Reality (VR) and Augmented Reality (AR) increases, the demand for kinesthetic haptic feedback devices is rapidly rising. Motor based haptic interfaces are heavy and bulky, leading to discomfort for the user. To address this issue, haptic gloves based on electrostatic clutches that offer fast response times and a thin form factor are being researched. However, high operating voltages and variable force control remain challenges to overcome. Electrostatic clutches utilizing functional polymers with charge accumulation properties and dielectric liquid can generate the frictional shear stress over a wide range from 0.35 N/cm² to 18.9 N/cm² at low voltages below 100 V. Based on this, the haptic glove generates a high blocking force and is comfortable to wear.

## 1 INTRODUCTION

As interest in Virtual Reality (VR) and Augmented Reality (AR) increases, the demand for haptic feedback devices is also rapidly rising. Haptic feedback is categorized into tactile and kinesthetic senses, kinesthetic sense is providing information about the shape and stiffness of objects. Traditional kinesthetic haptic gloves use motors to represent the form and rigidity of objects. [1] While this method offers easy control and fast response times, it has drawbacks such as low force density and discomfort for the user.

To address these issues, research on kinesthetic haptic gloves using soft actuators is being conducted. [2] Pneumatic actuators are widely used due to their strong output. However, hundreds of kPa of pressure is needed for presenting sufficient kinesthetic sensations. therefore, bulky and heavy pneumatic pumps are required. [3] This significantly diminishes the wearability of haptic gloves. Electrostatic clutches, which use electrostatic forces between two electrodes, are gaining attention in wearable haptic applications due to their thin form factor and powerful output. [4] Nevertheless, conventional electrostatic clutch-based haptic gloves require high input voltages ranging from hundreds to thousands of volts, resulting in bulky control circuits and potential interference with surrounding wearable devices. [5] Additionally, since the strip of the clutch is not perfectly flat, the electrodes and the dielectric only partially contact each other, and full contact occurs through a zipping propagation. As a result, the electrostatic clutch requires a high pull-in voltage.

To create a perfect electrostatic clutch based haptic glove, several key issues must be solved. 1) It must generate strong frictional forces while being thin enough to fit inside clothing. 2) It should offer a large range of frictional forces with a response time within tens of milliseconds. 3) The glove should be easy to don and doff. [6]

This paper proposes an electrostatic clutch that generates variable friction at operating voltages below 100 V and uses it to design a kinesthetic haptic glove. The electrostatic clutch uses a functional polymer with charge accumulation properties, which generates high electrostatic force even at low voltages.

Additionally, by using dielectric liquid to lower the pull-in voltage, a wide range of frictional forces from 0.35 N to 18.9 N can be achieved at voltages below 100 V. With a maximum thickness of 4 mm, it is thin enough to fit inside clothing and is convenient to don and doff.

## 2 ELECTROSTATIC CLUTCH

### 2.1 Electrostatic clutch with low voltage

A general electrostatic clutch is designed with a simple structure where a dielectric is placed between two flexible electrodes, providing frictional force induced by the electrostatic force. The simplest model of an electrostatic clutch is based on electrostatic attraction and dry Coulomb friction, which can be expressed by the following equation: [5]

$$F_{Friction} = \mu \times \frac{1}{2}\varepsilon_0\varepsilon_r AE^2$$

where $\mu$ is the friction coefficient, $\varepsilon_0$ the vacuum permittivity, $\varepsilon_r$ the relative permittivity of the dielectric.

In this paper, a clutch was designed using Polyvinyl chloride gel (PVC gel) with a high dielectric constant, achieving a high friction force. Figure 1(a) shows the dielectric constant of the PVC gel, measured at $2\times10^5$ at 0.001 Hz. The high dielectric constant of PVC gel is due to the negative charges injected at the cathode accumulating near the anode [7], displaying behavior different from conventional dielectrics where dipoles rotate within the electric field. Figure 1(b) illustrates the working principle of the PVC gel based electrostatic clutch.

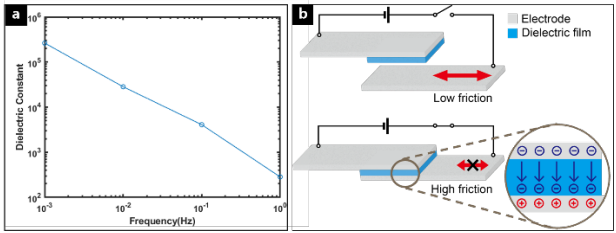

Fig.1 (a) Dielectric constant of PVC gel,
(b) Working principle of clutch

### 2.2 Friction tunable electrostatic clutch

In the case of a clutch with a simple structure where the dielectric is placed between two electrodes, the strip is not perfectly flat, resulting in a high pull-in voltage. To address this issue, high vertical pressure must be applied on the film. However, when full contact is achieved, the tackiness of the PVC gel causes high friction even without applying voltage.

To eliminate the tackiness of the PVC gel surface and increase the electrostatic force [8], a dielectric liquid was applied over the PVC gel. Figure 2 shows the voltage-frictional shear stress graph with and without the presence of dielectric liquid. With the dielectric liquid applied, the minimum operating voltage decreased from 40 V to 20 V, and the range of friction forces broadened from 0.35 N to 18.9 N. By using this clutch, we design a kinesthetic haptic glove that controls a wide range of blocking forces at low voltage levels.

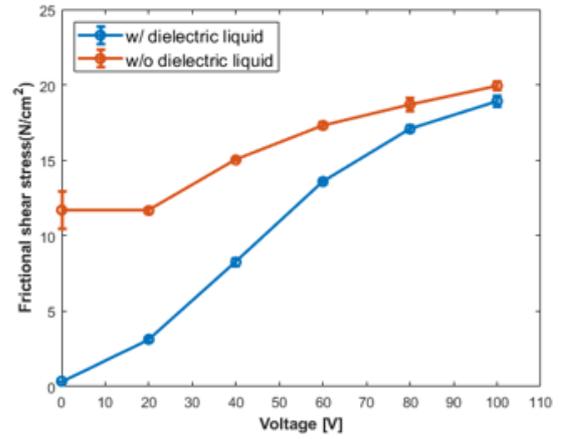

Fig.2 Performance of clutch

## 3 KINESTHETIC HAPTIC FEEDBACK GLOVE

Figure 3 shows the electrostatic clutch module and the haptic glove. The electrostatic clutch uses Al-coated PET and FPCB as electrodes, with a 50 μm thick PVC gel serving as the dielectric. To make initial contact, a cover was added, and a spring was attached to apply restorative force. The clutch module is very thin, with a thickness of 4 mm, making it easy to wear and capable of being attached inside clothing. The clutch module can offer various blocking forces depending on the applied voltage.

The Al-coated PET serves as the anode and functions as a tendon, effectively transmitting force to the fingers. Additionally, a guide is installed to ensure precise alignment. The electrostatic clutch is fixed to a wristband using Velcro, allowing for quick adjustment of the clutch module's position. This design enables the glove to adapt to different hand sizes.

Electrostatic force-based actuator, which operate at high driving voltages, require design considerations to enhance safety. Therefore, the operating voltage of the actuator was reduced to 100V to reduce potential risks, and the clutching area was enclosed with a cover to prevent contact with the electrodes.

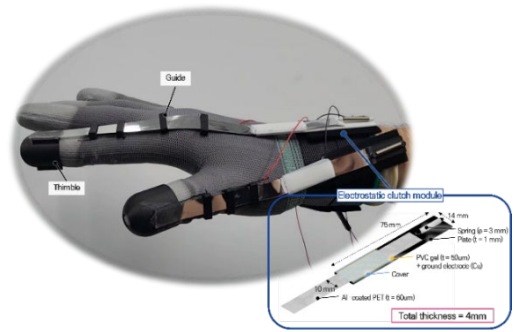

Fig.3 Kinesthetic haptic glove

## 4 Conclusion

PVC gel based electrostatic clutch provides a wide range of frictional forces at low voltages by using a charge accumulation properties of functional polymers. Based on this design, we propose a kinesthetic haptic glove. The clutch generates a high frictional shear stress of 18.9 N/cm² at a low voltage of 100 V, and it can control the frictional shear stress within the range of 0.35 N/cm² to 18.9 N/cm² by adjusting the voltage. Additionally, we designed a kinesthetic haptic glove using electrostatic clutch module with a thickness of 4 mm, providing strong resistance while ensuring comfortable wearability.

The clutch module, offering variable frictional forces at voltages below 100 V, features a compact electric circuit and minimal impact on surrounding wearable devices, making it highly advantageous and highly versatile for wearable applications. However, due to the high electrostatic force of the clutch, the aluminum coating on the PET was found to peel off, leading to a decline in performance as the number of operations increased. In the future, the deposition method of the electrodes will be improved to enhance the durability of the clutch.

## Acknowledgement

This work was supported by the National Research Council of Science & Technology (NST) grant by the Korea government (MSIT) (CRC23021-000)